\documentclass[12pt]{iopart}

\usepackage{iopams}  
\usepackage{pst-node}
\usepackage{graphicx}
\usepackage{dcolumn}
\usepackage{bm}
\usepackage{chemfig}
\usepackage{color}
\usepackage{siunitx}
\expandafter\let\csname equation*\endcsname\relax
\expandafter\let\csname endequation*\endcsname\relax
\usepackage{amsmath}
\usepackage{orcidlink}

\begin{document}

\title{Modeling Photothermal Effects in High Power Optical Resonators used for Coherent Levitation\\}

\author{Chenyue~Gu$^1$, Jiayi~Qin$^{1,2}$, Giovanni~Guccione$^1$, Jinyong~Ma$^3$, Ruvi~Lecamwasam$^{4,5}$
and Ping~Koy~Lam\orcidlink{0000-0002-4421-601X}$^{1,5}$}

\address{$^1$ \textit{ARC Centre of Excellence for Quantum Computation and Communication Technology (CQC$^2$T), Research School of Physics, The Australian National University, Canberra ACT 2601, Australia.}}
\address{$^2$ \textit{Centre for Gravitational Astrophysics (CGA), Research Schools of Physics \& Astronomy and Astrophysics, The Australian National University, Canberra ACT 2601, Australia.}}
\address{$^3$ \textit{ARC Centre of Excellence for Transformative Meta-Optical Systems, Research School of Physics, The Australian National University, Canberra ACT 2601, Australia.}}
\address{$^4$ \textit{Quantum Machines Unit, Okinawa Institute of Science and Technology Graduate University, Okinawa 904-0495, Japan.}}
\address{$^5$ \textit{A*STAR Quantum Innovation Centre (Q.InC), Institute for Materials Research and Engineering (IMRE), Agency for Science, Technology and Research (A*STAR), 2 Fusionopolis Way, 08-03 Innovis 138634, Republic of Singapore}}
\ead{ping.lam@anu.edu.au}

\vspace{10pt}

\begin{indented}
\item{\today}
\end{indented}


\begin{abstract}
    Radiation pressure can be used to enable optomechanical control and manipulation of the quantum state of a mechanical oscillator. Optomechanical interaction can also be mediated by photothermal effects which, although frequently overlooked, may compete with radiation pressure interaction. Understanding of how these phenomena affect the coherent exchange of information between optical and mechanical degrees of freedom is often underdeveloped, particularly in mesoscale high-power systems where photothermal effects can fully dominate the interaction. Here we report an effective theoretical model to predict and successfully reconstruct the dynamics of a unique optomechanical system: a cavity-enhanced setup for macroscopic optical levitation, where a free-standing mirror acts as the optomechanical oscillator. We decompose the photothermal interaction into two opposing light-induced effects, photothermal expansion, and thermo-optic effects. We then reconstruct a heuristic model that links the intracavity field to four types of cavity length changes caused by acoustic ($x_\textrm{ac}$), centre of mass ($x_\textrm{lev}$), photothermal ($x_\textrm{ex}$) and thermo-optic ($x_\textrm{re}$) displacements. This offers refined predictions with a higher degree of agreement with experimental results. Our work provides a mean to precisely model the photothermal effect of high power optomechanical systems, as well as for developing more precise photothermal modelling of photonics systems for precision sensing and quantum measurements.


\end{abstract}



\section{\label{sec:level1} Introduction}

Light fields carry momentum and can produce noticeable radiation pressure force~\cite{aspelmeyer2014cavity}. At high optical intensity, this can enable strong and controllable interactions between light and small objects~\cite{ashkin1980applications, gonzalez2021levitodynamics}. The high power density and photon flux of laser radiation, in the presence of loss, can lead to heating. For high-power laser applications, understanding the process dynamics caused by optical and thermal excitations is crucial for the control and stabilisation of these effects.~\par 

With exposure to any form of radiation, local heating can be caused on a sample via absorption and subsequent thermalization of energy. Among all laser-surface interactions, such \emph{photothermal} effects are a topic of fascination~\cite{hess1989photoacoustic}. They have found applications in spectroscopy of thin films~\cite{jackson1981photothermal, bertolotti1998analysis}, photothermal imaging~\cite{chien2020nanoelectromechanical, zhang2016depth, furstenberg2012chemical}, and as a feedback control or stabilisation tool~\cite{pinard2008quantum, qiu2022dissipative}. However, photothermal effects are unfavourable in optical systems because the optical absorption often causes thermal expansion and refractive index change. This induces measurement noise in sensing systems, such as the Laser Interferometer Gravitational-Wave Observatory (LIGO) system~\cite{zhao2006compensation, de2002experimental, evans2015observation, braginsky2001parametric, abbott2016gw150914}.~\par

In optomechanical systems, where light fields and mechanical systems are coupled via radiation pressure interaction, the study of photothermal effects is important. Optomechanical systems provide unique tools for applications where the highly sensitive measurement of small displacements, forces, masses, and accelerations are desired. They have found applications in high-precision sensors for atomic force microscopy~\cite{liu2012wide} and measurements at or beyond the standard quantum limit~\cite{kippenberg2007cavity, arcizet2006high, courty2003quantum, arcizet2006beating}. Recent studies have explored different physical realizations of cavity optomechanical systems employing cantilevers~\cite{kleckner2006sub, mow2008cooling}, micro-mirrors~\cite{arcizet2006radiation, gigan2006self}, micro-cavities~\cite{kippenberg2005analysis, schliesser2006radiation}, nano-membranes~\cite{thompson2008strong}, and macroscopic mirror modes~\cite{corbitt2007all}. The development of new and more extreme cavity systems at all scales, with increased optical densities and reduced mechanical oscillator sizes, has made it necessary to address photothermal effects. In such systems, the large mass of the oscillators can provide high inertial sensitivities resulting in lower tolerance to noise and mechanical oscillations. The interactions between optical field and mechanical oscillation induce parametrically amplified oscillations and nonlinear system dynamics. Photothermal effects mediate further exchanges between optical and mechanical degrees of freedom. In doing so they compete with --- and may even dominate over --- radiation pressure force~\cite{yoder2005opto}. This often causes parametric instabilities that inevitably affect the performance of the instruments~\cite{zhao2015parametric}. This is a common scenario in monolithic whispering gallery resonators~\cite{jiang2020optothermal, konthasinghe2017self}, micromechanical systems~\cite{rokhsari2005radiation}, and optical levitation configurations~\cite{ma2020observation}. Depending on the strength and relative timescale of each type of interaction, the consequences on the system can be surprisingly counter-intuitive and hard to predict. It is thus important to develop a simple and comprehensive model to predict the evolution of the systems in these scenarios.\par

Here we report a model for the dynamics of an optomechanical levitation system \cite{ma2020observation}. Due to the macroscopic mass of the levitation target, a very high power is needed. The system thus experiences strong non-linearities. Photothermal effects, competing with radiation pressure force, aggravate the parametric amplification in the system. Through theoretical modelling and analysis of experimental data, we show that it is necessary to decompose the photothermal expansion into two opposing effects with different timescales. This model can capture the essential features of the complex dynamics in a high-power optomechanical setup. The model will be useful for optomechanical systems where high-intensity lasers are used or high precision of the measurements is desired. It has wide-ranging implications for future quantum optomechanical systems and fundamental studies in quantum mechanics.~\par

\section{\label{sec:level2} Levitation experiment and model}

The model developed in this paper is of relevance to all the optical systems mentioned earlier. Here, however, we focus on one particular type that we use for testing and cross-validation of our results: a vertical cavity for optical levitation. In this system, the dynamics of the coupling of optical and mechanical degrees of freedom can be understood by considering the schematic of an optically driven Fabry-P\'erot cavity resonator, with one end mirror fixed and the other functioning as a mass-on-a-spring. Circulating power within the cavity exerts a force upon the ``movable'' mirror, changing the cavity length. The intracavity light field intensity and phase are then in turn modified by the new state of the cavity. This back-action mechanism causes the combined optical and mechanical system to reach a new equilibrium condition.

Optical levitation systems are interesting for their extreme environmental isolation, offering unique tools for precise metrology, optomechanical coupling, and preparation of mechanical quantum states. Among these systems, cavity levitation proposals such as those based on optical tripods~\cite{guccione2013scattering} or sandwich-like resonators~\cite{michimura2017optical} stand out for the particularly large target mass --- up to milligram-scale. Recent developments in levitating milligram-scale mirrors also suggest these systems as good candidates to explore quantum and nonlinear effects in the macroscopic regime~\cite{ma2020observation, guccione2013scattering, michimura2017optical, michimura2020quantum}. \par

\begin{figure*}\centering
    \includegraphics[width=0.99\textwidth]{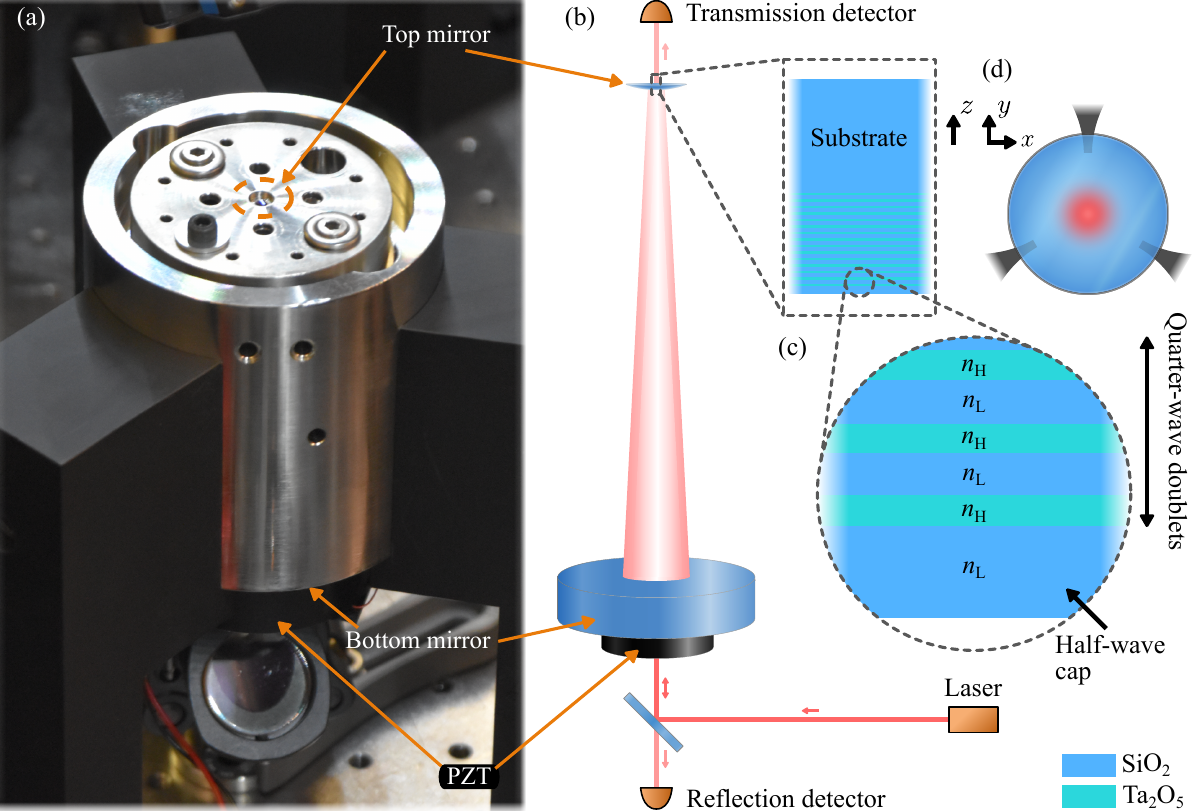}
    \caption{\label{fig:1} The levitation setup. (a) The vertical Invar cavity used in the levitation experiments. A small lightweight mirror is subject to the radiation pressure inside the cavity. The mirror is placed on the three small supporting points that are connected to the Invar case. The bottom mirror is attached to the piezoelectric actuator, which moves together to induce a linear scan of cavity detuning. (b) Schematic of the single laser levitation system. (c) High reflective mirror with $\textrm{Ta}_2 \textrm{O}_5$ and $\textrm{Si} \textrm{O}_2$ bi-layers used as a distributed Bragg reflector. (d) Levitation mirror placed on the Invar mount.}
\end{figure*}


The system under consideration in this paper is inspired by a simplified version of these optical levitation cavities. It consists of a vertical Fabry-P\'erot cavity, where the top end mirror is free-standing, unclamped, as a free mass at the top of the cavity, as shown in the FIG.~\ref{fig:1}(a). The top mirror of the cavity is coated with \chemfig{Ta_2 O_5} and \chemfig{Si O_2} bi-layers acting as a distributed Bragg reflector (presented in FIG.~\ref{fig:1}(c)), which provides a reflectivity as high as 99.992\% on the curved side of a fused silica spherical cap. The mass of the levitated mirror studied here is $1.116\pm0.003$ \si{\mg} with a radius of curvature of \SI{25}{\milli\metre}, diameter of \SI{3}{\milli\metre}, and thickness of around \SI{50}{\micro\metre}. The error on the reported value of the mirror’s weight was given by the accuracy of a high-precision scale from the ANU School of Chemistry. The optical power required to levitate a mass at this scale is high enough to deform the mirror's surface. Due to the relatively small spot size of the laser, the intensity at the surface of the mirror can reach as high as \SI{3}{\mega\watt\per\centi\metre\squared}, which is even larger than the intensity in the baseline LIGO interferometers~\cite{abbott2016observation}, resulting in significant photothermal effects. The input light is introduced via the bottom mirror, which is a conventional \ang{;;1} concave mirror, with a high-reflectivity coating (99.9\%) on a fused silica substrate and mounted on a piezoelectric actuator to control the cavity length. The whole cavity is approximately \SI{80}{\milli\metre} long and is enclosed by a monolithic Invar case to reduce thermal fluctuations (FIG.~\ref{fig:1}(a)). A spacer on the top sketched in FIG~\ref{fig:1}(d), which is fixed to the Invar case, has a hole with three small symmetric contact points to support the mirror when it is not lifted or levitated. This is designed to minimise Van der Waals forces. A \SI{1050}{\nano\metre} Nd:YAG laser is coupled into the cavity via the bottom mirror, producing an input light with up to \SI{20}{\watt} of power, which can generate an intracavity power of around \SI{11.5}{\kilo\watt} with a finesse of 2850.\par 

When the radiation pressure exerted by the intracavity field is sufficiently strong to overcome the weight of the mirror, the mirror is lifted off one of the contact points and thus the cavity length is increased. Under these conditions the strength of the optical spring on the mirror changes. The total of the radiation pressure and gravitational forces of the mirror provides a restoring force, creating the optical trap for the macroscopic mechanical oscillator in this levitation setup~\cite{guccione2013scattering}. By moving the bottom mirror with the piezoelectric actuator, we change the resonance condition of the cavity, effectively changing the cavity detuning while keeping the laser frequency constant. To monitor the evolution of the intracavity field, we measure the reflected and the transmitted outputs of the cavity.\par

In this optomechanical system, thermal effects can also modify the system dynamics. There are several ways in which photothermal effects can influence the cavity~\cite{konthasinghe2017self}. One is by thermal expansion: by absorbing optical power, the coating and substrate of the mirrors can expand and cause a reduction in cavity length. Another possibility is the thermo-optic effect, where the refractive index of a material is modified due to a variation in temperature, thus changing the effective optical path of the cavity. This effect is not negligible, and we will see in fact that it plays a significant role in this system.\par 

A simple model to describe the effective photothermal displacement in a cavity assumes a relaxation process linearly proportional to the optical power~\cite{dhurandhar1997stability, marino2006canard}. This empirical law originates from a single-pole approximation of the full thermo-elastic response of the mirror~\cite{cerdonio2001thermoelastic} in the case of photothermal expansion. Despite its simplicity, this equation proved its success in predicting the complicated dynamical behaviour of optomechanical cavities under the regime where the interaction between intracavity field and the mirror is weak~\cite{ma2020observation}.\par

In the current model, the modification of the effective optical cavity length is separated into three different entities: acoustic ($x_{\rm ac}$), centre-of-mass ($x_{\rm lev}$) and photothermal ($x_{\rm th}$) displacements. The acoustic motion evolves as a harmonic oscillator, the change of the centre of mass is due to the push of radiation pressure force, and the thermal motion is induced by photothermal effects. The equations of motion for the system in the rotating frame can then be written as
\begin{eqnarray}
        \label{eq:1}
        \dot{a}&&=
        \left[-\kappa+i(\Delta+G(x_\textrm{lev}+x_\textrm{th}+x_\textrm{ac}))\right]a+\sqrt{2\kappa_\textrm{in}}a_\textrm{in},\\
        \label{eq:2}
        \dot{x}_\textrm{th}&&=-\gamma_\textrm{th}(x_\textrm{th}+\beta_\textrm{th} P_\textrm{opt}(a)),\\
        \label{eq:3}
        \ddot{x}_\textrm{ac}&&= -\gamma_\textrm{ac}\dot{x}_\textrm{ac}-\omega_\textrm{ac}^2x_\textrm{ac}+F_\textrm{opt}(a)/m_\textrm{ac},\\
        \label{eq:4}
        \ddot{x}_\textrm{lev}&&=
        \begin{cases}
            -\gamma_\textrm{lev}\dot{x}_\textrm{lev}, & \textrm{when supported}, \\
            -\gamma_\textrm{lev}\dot{x}_\textrm{lev}+(F_\textrm{opt}(a)-F_\textrm{g})/m, & \textrm{when not supported}.
        \end{cases}
\end{eqnarray}
These dynamic equations characterise the experimental system in the classical limit. The evolution of the optical field is shown as Eq.~(\ref{eq:1}), where $a$ denotes the amplitude of the intracavity field. The amplitude of the driving field is $a_\textrm{in}$, coupling through the input mirror at a rate $\kappa_\textrm{in}$. The imaginary part is the oscillatory term (in the rotating frame of the cavity this is described by the detuning $\Delta=\omega_\textrm{l}-\omega_\textrm{opt}$, where $\omega_\textrm{l}$ and $\omega_\textrm{opt}$ are angular frequencies of the optical field and the cavity resonance respectively). The total loss is described as $\kappa$ (with $\kappa \ge \kappa_\textrm{in}$), which defines the decay rate of the cavity. The evolution of the amplitude of the intracavity field $a$ depends on a shift of the centre of the mass $x_\textrm{lev}$ caused by the radiation pressure force, the displacement of the mirror's surface $x_\textrm{ac}$ following vibrations of acoustic mode, and difference of the cavity length $x_\textrm{th}$ due to total photothermal effects. The coefficient $G=\omega_\textrm{opt}/L_\textrm{cav}$ is the optomechanical coupling strength between the mirror and the intracavity field, where $L_\textrm{cav}$ is the length of the cavity. Equation~(\ref{eq:2}) gives the empirical model of photothermal effects. The thermal relaxation rate $\gamma_\textrm{th}$ and the susceptivity coefficient $\beta_\textrm{th}$ are dependent on the properties of the mirror's material and on the absorption coefficient. The evolution of the centre of mass of the mirror is described in Eq.~(\ref{eq:4}). The term $F_\textrm{opt}=\hbar G|a|^2=2P_\textrm{opt}/c$ is radiation pressure force, calculated from the intracavity power $P_\textrm{opt}$ and the speed of light $c$. The gravitational weight of the mirror is denoted as $F_\textrm{g}=mg$, where $g$ is the free-fall gravitational acceleration and $m$ is the total inertial mass of the mirror. The dissipation coefficient of the centre-of-mass motion is described by $\gamma_\textrm{lev}$. When there is sufficient power in the cavity, the radiation pressure force can lift the mirror above its supporting stand. The net force acting on the mirror thus is given by the balance of radiation pressure and the mirror's weight. Otherwise, when the intracavity power is not strong enough to lift the mirror, the net force is 0. The forces on the mirror may also excite the natural vibrational modes of the mirror. These are described by an ``acoustic'' degree of freedom, which behaves as a harmonic oscillator of frequency $\omega_\textrm{ac}$ and damping $\gamma_\textrm{ac}$, described in Eq.~(\ref{eq:3}). The effective mass of this acoustic mode, $m_\textrm{ac}$, is generally a fraction of the total inertial mass of the mirror ($m_\textrm{ac} < m$).\par

In section~\ref{sec:level4}, we will show that this model provides a faithful picture of the dynamics of the levitation system in the regime of weak interaction, where lower input power or faster scanning of cavity detuning is applied and when parametric amplification of the oscillation is not observed in the transmission. In the scenario of higher input power, when the amplified oscillation is involved, as we will see in section~\ref{sec:level4}, the model becomes less faithful due to the ineffective fitting for the features of the parametric amplification of the oscillation in experiments. This is mainly caused by the failure of modelling the thermal effects which introduce, and further enhance, the parametric gain that can destabilise the system. This emphasises the importance of a more detailed investigation on photothermal effects. In the next section, we will focus on two different photothermal effects that influence the cavity, and modify the equations of motion of the system by introducing two photothermal displacements instead of only one.\par

\section{\label{sec:level3} Refining the model}

With an extremely high intracavity field of a few kilowatts, and a small beam diameter on the coating of around \SI{100}{\micro\metre}, even a small fraction of radiation absorption can cause a local rise in temperature in the coating. This results in photothermal expansion and thermo-optic effects that modify the effective cavity length.\par

\begin{figure}\centering
    \includegraphics[width=0.5\textwidth]{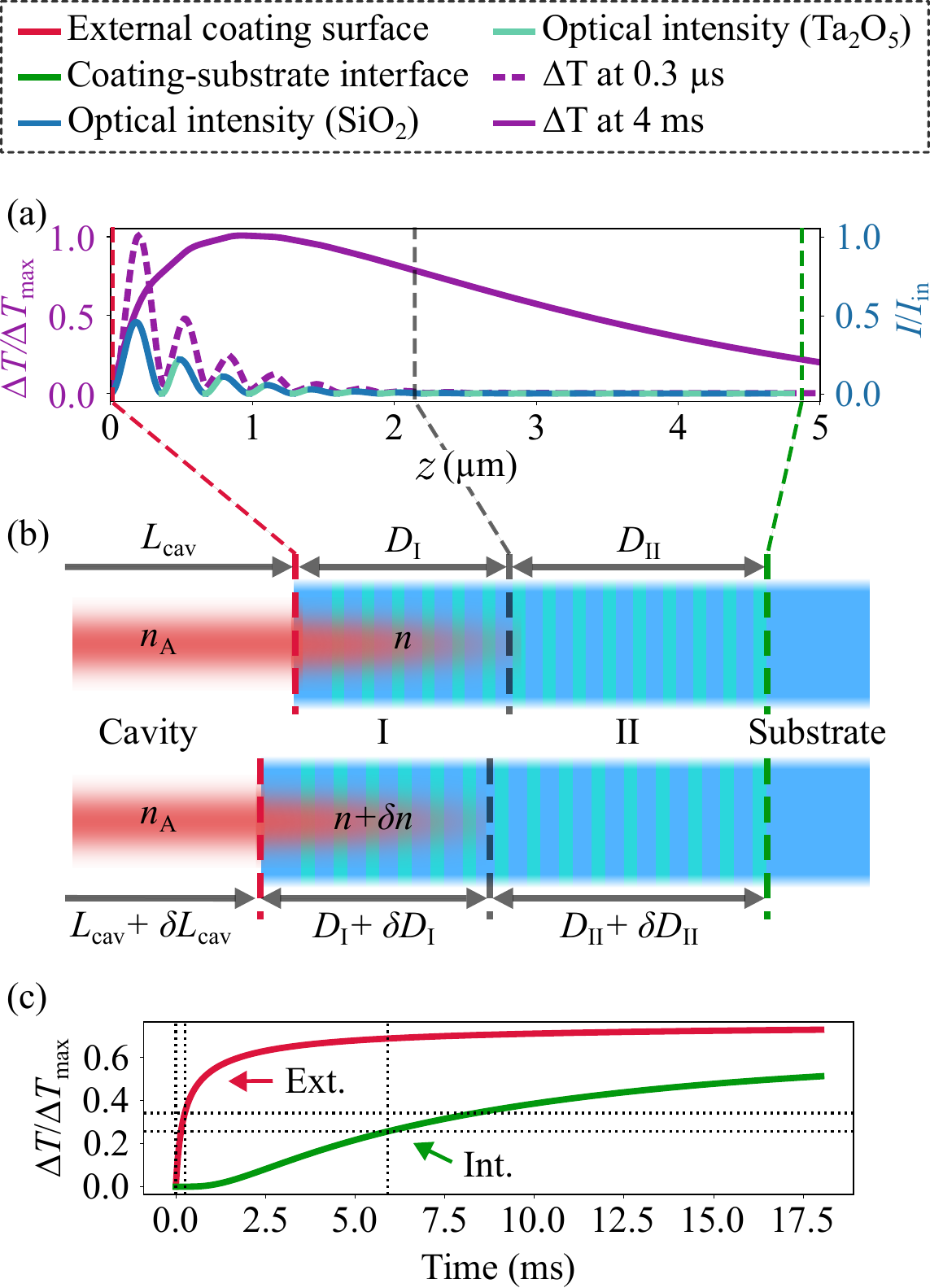}
    \caption{\label{fig:2} Phothermal expansion and thermo-optic effects in the dielectric Bragg coating of the top mirror. The darker and lighter blue represent the light field inside the layer of $\textrm{SiO}_\textrm{2}$ and $\textrm{Ta}_\textrm{2}\textrm{O}_\textrm{5}$ respectively. The red and green dashed lines that connect (a) and (b) represent the external coating surface and the coating-substrate interface. (a) Simulation of the optical intensity of the light field inside the cavity. The dashed purple curve gives the temperature change at \SI{0.3}{\micro\second} after the laser shone on the mirror. The solid purple line shows the thermal distribution inside the levitation mirror coating at \SI{4}{\milli\second}. (b) Schematic of the expansion and refractive index change of the top mirror due to photothermal effects, which cause effective optical length variation. The upper figure illustrates the coating before it absorbs the heat from intracavity field and the lower figure shows the material after response to the photothermal effects. The gray dashed line divides the coating into part (\uppercase\expandafter{\romannumeral1}) and (\uppercase\expandafter{\romannumeral2}) where 1\% light transmits through. (c) Simulation of the average temperature changes on external coating surface (\SI{1}{\micro\metre} thick, plotted in red) and coating-substrate interface (\SI{0.5}{\micro\metre} thick coating with \SI{0.5}{\micro\metre} thick substrate, plotted in green).}
\end{figure}

To undertake an in-detail analysis on these two effects, we first calculate the optical intensity inside the mirror coating. The top mirror is coated with a dielectric Bragg coating designed for high reflectivity. The coating is made of quarter-wave doublets and a half-wave cap layer, shown in FIG.~\ref{fig:2}(b) with the low-index material \chemfig{Ta_2 O_5} and the high-index material \chemfig{Si O_2}. These two materials are common choices for our wavelength of \SI{1064}{\nano\metre}, because of a high difference in refractive index and very low absorption at the wavelength under consideration. In FIG.~\ref{fig:2}(a), we give a simulation of the intensity of the light field. The dashed grey line is drawn at the layer around \SI{2}{\micro\metre} deep in the coating, where 99\% of the light is reflected. Using this as a boundary, we divide the top mirror coating into two different parts, shown in FIG.~\ref{fig:2}(a--b): part (\uppercase\expandafter{\romannumeral1}) close to the surface (red dashed line) with thickness $D_\textrm{\uppercase\expandafter{\romannumeral1}}$ and part (\uppercase\expandafter{\romannumeral2}) connected to substrate (green dashed line) with thickness  $D_\textrm{\uppercase\expandafter{\romannumeral2}}$. The laser shone on the mirror not only produces optical energy but also thermal energy. The heat from the intracavity field is thus absorbed by the mirror coating and then transferred into the deeper layers, causing a temperature gradient inside the coating along the $z$ axis.\par

In order to investigate how the mirror responds to the intracavity field by absorbing the thermal energy from it, we estimated the thermal conduction of the coating by numerically solving the heat diffusion equation:
\begin{eqnarray}
     \label{eq:17}
     \alpha\nabla ^{2}T(x,y,z,t)-\frac{\partial T(x,y,z,t)}{\partial t}=0,
\end{eqnarray}
where $T(x,y,z,t)$ denotes temperature. We use a cylindrical mirror with a coating formed of two materials, \chemfig{Ta_2 O_5} and \chemfig{Si O_2}, as shown in Figure.~\ref{fig:2}(b). The thermal diffusivity $\alpha=\kappa_\textrm{th} / s\rho$ of \chemfig{Ta_2 O_5} is $1.6\times 10^{-8}$ \si{\metre\squared\per\second}, where $s=5.0\times10^5$ \si{\joule\per\kilo\gram\per\kelvin} is the specific heat, $\rho=4.2\times10^3$ \si{\kilo\gram\per\cubic\metre} is the density and $\kappa_\textrm{th}=33$ \si{\watt\per\metre\per\kelvin} is the thermal conductivity. For \chemfig{Si O_2}, the thermal diffusivity $\alpha$ is $8.4\times 10^{-10}$ \si{\metre\squared\per\second}, where $s=7.45\times10^5$ \si{\joule\per\kilo\gram\per\kelvin}, $\rho=2.2\times10^3$ \si{\kilo\gram\per\cubic\metre} and $\kappa_\textrm{th}=1.38$ \si{\watt\per\metre\per\kelvin}~\cite{evans2008thermo}. We note that we are extracting only qualitative features from this simulation, which will not be affected by slight variations in the values of these parameters.  Since the cavity power decays sharply inside the coating, the heat is deposited primarily near the surface. We assume that the mirror experiences a source of thermal heat with spatial distribution proportional to the optical intensity within the mirror coating:
\begin{eqnarray}
     \label{Eq:Gaussian_flux}
     \frac{\partial T(x,y,z,t)}{\partial t}=D(z)\frac{P_\textrm{circ}\mathcal{A}}{\rho s \pi w^2_\textrm{0}}e^{-2\frac{x^2+y^2}{w^2_\textrm{0}}},
\end{eqnarray}
where $D(z)$ is the optical distribution as a function of $z$, which is given as the blue line in FIG.~\ref{fig:2}(a). The waist size $w_\textrm{0}$ on the surface of the mirror is around \SI{100}{\micro\metre}, according to the resonant cavity mode, which is constant in time, at its surface center. We estimate the heat transfer with an ideal absorption coefficient of $\mathcal{A}=10~\textrm{ppm}$ \si{\per\cm} and an input power of \SI{1}{\watt} that gives a power circulating in the cavity of $P_\textrm{circ}\approx\SI{300}{\watt}$. On the inner side of the substrate, after the medium of diffusion switches to the fused silica substrate, we homogenize the temperature flux with Robin boundary conditions:
\begin{eqnarray}
     \label{Eq:Robin_boundary_condition}
     \kappa_\textrm{th} \vec{n}\cdot\vec{\nabla}T(x,y,z_\textrm{l},t)+h\left(T-T_\textrm{ext}\right)=0,
\end{eqnarray}
where $h=1.65\times10^{8}$ \si{\watt\per\metre\squared\per\kelvin} is the thermal contact conductance which is defined as the reciprocal of thermal contact resistance~\cite{madhusudana2014thermal}. $T(x,y,z_\textrm{l},t)$ represents the temperature at the end of the material simulated, which is \SI{5}{\micro\metre} inside the substrate. The term $T_\textrm{ext}$ denotes the temperature of the environment, which we assume is constant.\par

Figure~\ref{fig:2}(c) reveals that the temperature at the external surface rises fast in the beginning, and changes slowly after \SI{2}{\milli\second} have elapsed. By contrast, the temperature at the interface of the substrate with the coating that is around \SI{5}{\micro\meter} inside the coating takes around \SI{1}{\milli\second} to start changing. Since the thermo-optic effect is primarily concentrated at the surface of the coating ($D_\textrm{\uppercase\expandafter{\romannumeral1}}$), while the thermal expansion contributes over the entire coating ($D_\textrm{\uppercase\expandafter{\romannumeral1}}$ and $D_\textrm{\uppercase\expandafter{\romannumeral2}}$), we conclude that the timescale for the two processes of thermal expansion and thermo-optic effects differs~\cite{konthasinghe2017self}. The surface layers of the coating thus experience ``fast'' refractive index variation, while the deeper layers experience ``slow'' expansion from thermal energy. This gives two different relaxation rates for the different thermal effects, emphasizing the importance of separating them into two independent terms. In FIG.~\ref{fig:2}, the simulation of the temperature change at \SI{0.3}{\micro\second} from the heat source acting on the mirror, and the thermal distribution at \SI{4}{\milli\second} are illustrated in dashed and solid purple lines. Here, we plot the thermal distribution at \SI{4}{\milli\second} because all our resonance scans in the experiment are faster than \SI{4}{\milli\second}. Note that the absorption of the mirror coating in the experimental setup is usually not ideal and the system is sometimes operated under high input power (as high as \SI{5}{\watt}). Both higher input power and larger absorption will cause more heat to transfer into the mirror coating, hence resulting in higher temperature rise. From Eq.~(\ref{Eq:Gaussian_flux}), however, we know that the values of the intracavity power $P_\textrm{circ}$ and the absorption $\mathcal{A}$ don't change the evolution of the thermal conduction. The time taken to reach equilibrium (FIG.~\ref{fig:2}(c)) and the distribution of the temperature (FIG.~\ref{fig:2}(a)) thus remain the same with different $P_\textrm{circ}$ and $\mathcal{A}$.\par

As we mentioned earlier, temperature change can cause expansion and changes to the refractive index of the mirror material. Our simulation in FIG.~\ref{fig:2} tells us that the temperature change of the substrate has less impact on the expansion of the mirror’s substrate, and the light is almost fully reflected from the interface between part (\uppercase\expandafter{\romannumeral1}) and (\uppercase\expandafter{\romannumeral2}). Thus, we ignore the expansion of the substrate in the following derivation. Under this estimation, we deduce that part (\uppercase\expandafter{\romannumeral1}) responds to the temperature change on both expansion and refractive index change, whereas in part (\uppercase\expandafter{\romannumeral2}) only the thermal expansion effect is considered since very little optical power penetrates through to experience a change in index. To first-order approximation, the effective cavity lengths before (cold) and after (hot) the top mirror absorbs the thermal energy of the optical field are given by:
\begin{eqnarray}
    \label{eq:5}
    &x_\textrm{cold}=L_\textrm{cav}n_\textrm{A}+D_\textrm{\uppercase\expandafter{\romannumeral1}}n,&\\
    \label{eq:6}
    &x_\textrm{hot}=(L_\textrm{cav}+\delta L_\textrm{cav})n_\textrm{A}+(D_\textrm{\uppercase\expandafter{\romannumeral1}}+\delta D_\textrm{\uppercase\expandafter{\romannumeral1}})(n+\delta n).
\end{eqnarray}\par

Equation~(\ref{eq:5}) gives the effective optical length of the cavity which consists of the effective length outside and inside the mirror coating before the mirror's surface reacts to the thermal effects. The refractive index of the medium of the cavity is $n_\textrm{A}$, which is air or vacuum in our system. We assume thermal changes to this refractive index to be negligible, if not zero (in vacuum). For a demonstrative calculation, it suffices to consider only the average refractive index of the coating $n$ in the following derivation of the two independent thermal effects. With thermal absorption, the expansion of the coating causes an increase of $\delta D_\textrm{\uppercase\expandafter{\romannumeral1}}$ in the thickness of part (\uppercase\expandafter{\romannumeral1}) and $\delta D_\textrm{ \uppercase\expandafter{\romannumeral2}}$ in part (\uppercase\expandafter{\romannumeral2}), which results in a reduction of the distance between two mirrors $\delta L_\textrm{cav}=-(\delta D_\textrm{\uppercase\expandafter{\romannumeral1}}+\delta D_\textrm{\uppercase\expandafter{\romannumeral2}})$. The schematic diagram of the change of the coating is illustrated in FIG.~\ref{fig:2}(b) where we only note the change, $n+\delta n$, in the average refractive index of part (\uppercase\expandafter{\romannumeral1}), which will influence the effective optical length of the cavity. Thus the effective optical length can be written as Eq.~(\ref{eq:6}). The substrate is considered to act as a thermal reservoir with direct radiative thermal energy exchanged with the environment.\par

Ignoring higher order terms, the variation of the effective optical length of the cavity due to thermal effects is:
\begin{eqnarray}
    x_\textrm{th}(D,n)&=&x_\textrm{cold}-x_\textrm{hot},\nonumber\\
    \label{eq:8}
    &=&-(\delta D_\textrm{\uppercase\expandafter{\romannumeral1}}+\delta D_\textrm{\uppercase\expandafter{\romannumeral2}}) n_\textrm{A}+\delta (D_\textrm{\uppercase\expandafter{\romannumeral1}}n).
\end{eqnarray}
Equation~(\ref{eq:8}) gives a clearer perspective of the system's reaction to the thermal effects: the first term indicates the reduction of the effective optical length from the decrease of the distance between the input and output mirrors of the cavity, while the second term represents the effective optical length change \emph{inside} the levitated mirror. By writing it as
\begin{eqnarray}
    \label{eq:9}
    x_\textrm{th}(D,n)=-(n_\textrm{A}-n)\delta D_\textrm{\uppercase\expandafter{\romannumeral1}}-\delta D_\textrm{\uppercase\expandafter{\romannumeral2}}n_\textrm{A}+D_\textrm{\uppercase\expandafter{\romannumeral1}}\delta n,
\end{eqnarray}
we can separate the thermal expansion effect on effective optical length change in the coating part (\uppercase\expandafter{\romannumeral1}) ($x_\textrm{ex\uppercase\expandafter{\romannumeral1}}$) and (\uppercase\expandafter{\romannumeral2}) ($x_\textrm{ex\uppercase\expandafter{\romannumeral2}}$) from the thermo-optic effect on effective optical length change ($x_\textrm{re}$) caused by the variation of refractive index:
\begin{eqnarray}
    \label{eq:10}
    x_\textrm{ex\uppercase\expandafter{\romannumeral1}}(D_\textrm{\uppercase\expandafter{\romannumeral1}}(T))&&=-(n_\textrm{A}-n)\mathrm{d} D_\textrm{\uppercase\expandafter{\romannumeral1}}, \\
    \label{eq:11}
    x_\textrm{ex\uppercase\expandafter{\romannumeral2}}(D_\textrm{\uppercase\expandafter{\romannumeral2}}(T))&&=-n_\textrm{A}\mathrm{d} D_\textrm{\uppercase\expandafter{\romannumeral2}},\\
    \label{eq:12}
    x_\textrm{re}(n(T))&&=D_\textrm{\uppercase\expandafter{\romannumeral1}}\mathrm{d} n.
    \end{eqnarray}
We assume that the mirror's expansion term $D_\textrm{\uppercase\expandafter{\romannumeral1}}$ and $D_\textrm{\uppercase\expandafter{\romannumeral2}}$ and the (average) refractive index of the coating materials $n$ depend only on local temperature $T$. This increases due to absorption of optical energy and cools as heat is conducted into the rest of the substrate.

We can calculate the time derivative of the effective optical length change due to expansion on coating part (\uppercase\expandafter{\romannumeral1}) as (detailed discussion can be found in~\ref{sec:A1}):
\begin{eqnarray}
    \label{eq:13}
    \dot{x}_\textrm{ex\uppercase\expandafter{\romannumeral1}}
    &=&-\gamma_\textrm{ex}(x_\textrm{ex\uppercase\expandafter{\romannumeral1}}+\beta_\textrm{ex\uppercase\expandafter{\romannumeral1}}P_\textrm{opt}),
\end{eqnarray}
where $\gamma_\textrm{ex}$ and $\beta_\textrm{ex\uppercase\expandafter{\romannumeral1}}=\partial D_\textrm{\uppercase\expandafter{\romannumeral1}}/\partial_\textrm{opt}$ are, respectively, the thermal relaxation rate and susceptivity coefficient of photothermal expansion. 
An analogous calculation also gives the derivative of the effective optical length change of part (\uppercase\expandafter{\romannumeral2}):
\begin{eqnarray}
    \label{eq:14}
    \dot{x}_\textrm{ex\uppercase\expandafter{\romannumeral2}}=-\gamma_\textrm{ex}(x_\textrm{ex\uppercase\expandafter{\romannumeral2}}+\beta_\textrm{ex\uppercase\expandafter{\romannumeral2}}P_\textrm{opt}),
\end{eqnarray}
where $\beta_\textrm{ex\uppercase\expandafter{\romannumeral2}}=\partial D_\textrm{\uppercase\expandafter{\romannumeral2}}/\partial P_\textrm{opt}$.\par

Here, the thermal expansion effect in part (\uppercase\expandafter{\romannumeral1}) and (\uppercase\expandafter{\romannumeral2}) share the same value of $\gamma_\textrm{ex}$ because it only depends on the coating material. Therefore, we can combine the thermal expansion of the two parts of the coating in Eq.~(\ref{eq:13}) and Eq.~(\ref{eq:14}), and rewrite them as:
\begin{eqnarray}
    \label{eq:15}
    \dot{x}_\textrm{ex}=-\gamma_\textrm{ex}(x_\textrm{ex}+\beta_\textrm{ex}P_\textrm{opt}),
\end{eqnarray}
where $\dot{x}_\textrm{ex}=\dot{x}_\textrm{ex\uppercase\expandafter{\romannumeral1}}+\dot{x}_\textrm{ex\uppercase\expandafter{\romannumeral2}}$, $x_\textrm{ex}=x_\textrm{ex\uppercase\expandafter{\romannumeral1}}+x_\textrm{ex\uppercase\expandafter{\romannumeral2}}$ and $\beta_\textrm{ex}=\beta_\textrm{ex\uppercase\expandafter{\romannumeral1}}+\beta_\textrm{ex\uppercase\expandafter{\romannumeral2}}=\partial D/\partial P_\textrm{opt}$.\par

Similarly, we can get the derivative of the effective optical length variation of thermo-optic effects as
\begin{eqnarray}
    \label{eq:16}
    \dot{x}_\textrm{re}&=&\frac{\mathrm{d}n(T(\varepsilon ,t))}{\mathrm{d}t}D_\textrm{\uppercase\expandafter{\romannumeral1}}\nonumber\\
    &=&-\gamma_\textrm{re}(x_\textrm{re}-\beta_\textrm{re}P_\textrm{opt}),
\end{eqnarray}
where $\gamma_\textrm{re}$ and $\beta_\textrm{re}=\partial n/\partial P_\textrm{opt}$ are the thermal relaxation rate and susceptivity coefficient of thermo-optic effects respectively. It is worthwhile to emphasise the different sign before the susceptivity coefficient compared to thermal expansion, which intuitively, is reasonable. Because when the refractive index of the coating increases, the effective optical path length inside the coating increases, which increases the effective cavity length. This indicates that with the positive $\beta_\textrm{re}$, the refractive index increases when absorbing thermal energy, reducing the deduction of effective optical length. Whereas, in our system, with a negative $\beta_\textrm{re}$, the refractive index decreases after absorbing thermal energy, leading to a further decrease of the effective optical length. \par

Replacing Eq.~(\ref{eq:2}) in the system of equations of the levitation cavity by the two independent equations, Eq.~(\ref{eq:15}) and Eq.~(\ref{eq:16}), together with Eq.~(\ref{eq:3}) and Eq.~(\ref{eq:4}), we derive the updated model containing four displacement degrees of freedom interacting with the intracavity field: photothermal expansion effect, thermo-optic effect, excitation of acoustic vibrations, partial lift-off of the mirror due to radiation pressure force. Thus, Eq.~(\ref{eq:1}) becomes:
\begin{eqnarray}
        \label{eq:20}
        \dot{a}=&&
        \left[-\kappa+i(\Delta+G(x_\textrm{lev}+x_\textrm{ex}+x_\textrm{re}+x_\textrm{ac}))\right]a\nonumber\\
        &&+\sqrt{2\kappa_\textrm{in}}a_\textrm{in},
\end{eqnarray}
and for the refined model the system of equations are Eqs.~(\ref{eq:3}--\ref{eq:4}) and (\ref{eq:15}--\ref{eq:20}).\par

\section{\label{sec:level4} Analysis and discussion}

\begin{figure}\centering
    \includegraphics[width=0.7\textwidth]{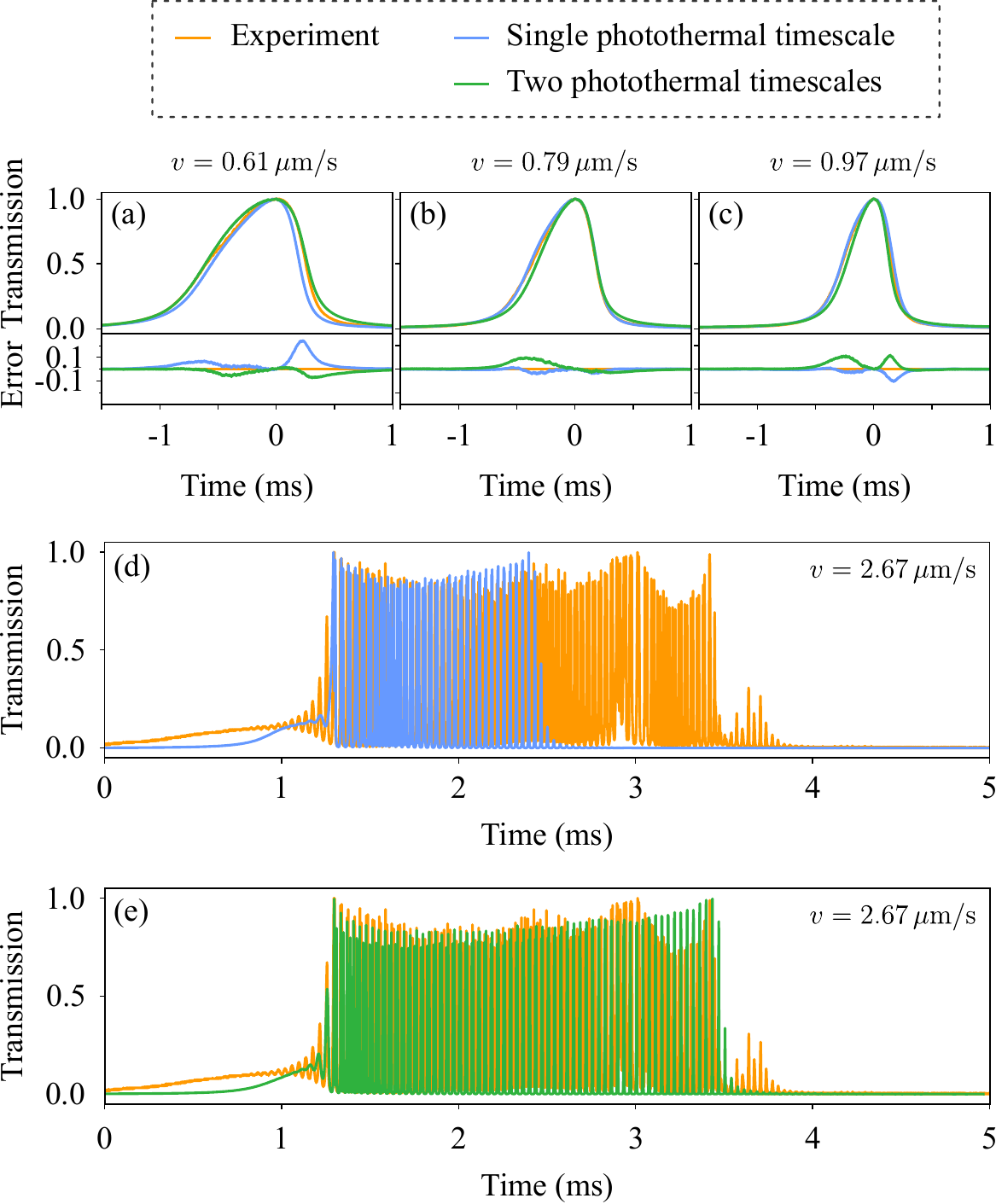}
    \caption{\label{fig:3} Time series of the optical output of the system driven by low (a--c) and high (d--e) input power. (a--c) Traces of the cavity response obtained through the transmission output for an input power of \SI{93}{\milli\watt}. Experimental data is represented by the orange line, whereas the simulations of the model with single photothermal-effect term are in blue while the simulations from the refined model are in green. Error is calculated with the ratio of the difference between simulation and experimental data to the experimental data. (d) and (e) give the time simulation results of the models with single and two photothermal timescales with an input power of \SI{2.75}{\watt} and a scan speed of \SI{2.67}{\micro\metre\per\second}. Note that this power is still below the threshold of levitating the mirror. Parametric amplification and complex evolution arise at high power. Compared with the simulation of the model with a single photothermal effect (blue), the simulation from the refined model (green) generally agrees with the experimental results (orange). The cavity loss used in simulation gets from experiments, which for (a--c) is $\kappa = 2\pi\times\SI{658}{\kilo\hertz}$ and for (d--e) is $\kappa = 2\pi\times\SI{735}{\kilo\hertz}$. Parameters inserted in the two different models are, for the model with single photothermal timescale, $\gamma_\textrm{th}=2\pi\times\SI{450}{\hertz}$ and $\beta_\textrm{th}=\SI{9}{\pico\metre\per\watt}$, for the model with two photothermal timescales, $\gamma_\textrm{ex}=2\pi\times\SI{11}{\hertz}$, $\beta_\textrm{ex}=\SI{120}{\pico\metre\per\watt}$, $\gamma_\textrm{re}=2\pi\times \SI{4000}{\hertz}$, and $\beta_\textrm{re}=\SI{-2}{\pico\metre\per\watt}$. }
\end{figure}

The simulation of the original model~\cite{ma2020observation} and the refined model for the cavity driven by low and high input power are shown in FIG.~\ref{fig:3}, which compares experimental measurements of the normalized intensity of cavity transmission with the simulated results. The cavity is measured by a linear red-to-blue (lower- to higher-frequency) detuning scan. The data is considered both at low (FIG.~\ref{fig:3}(a--c)) and high (FIG.~\ref{fig:3}(d--e)) power to compare different regimes. At low power, traces are aligned at their maximum, while at high power, traces are aligned at the point when the transmission first reaches the maximum.\par 

In FIG.~\ref{fig:3}(a--c), from left to right, we present the system under different scan speeds from \SI{0.61}{\micro\metre\per\second} to \SI{0.97}{\micro\metre\per\second}, with a low input power of \SI{93}{\milli\watt}. The difference between the experimental and the simulation results are shown as error in the figure. Comparing the three traces with different scan speeds, it is easy to see that with higher speed (FIG.~\ref{fig:3}(c)) the transmission has a more Lorentzian-shaped resonance. With lower scan speed (FIG.~\ref{fig:3}(a)), however, the traces are broadened, because photothermal effects blue-detune the resonance proportionally to the intra-cavity power, therefore dragging the maximum cavity output forward in the scan~\cite{qin2022cancellation}. These features are well simulated by the model with single photothermal timescale (Eqs.~(\ref{eq:1}--\ref{eq:4})) and the refined models with two photothermal effects (Eqs.~(\ref{eq:3}--\ref{eq:4}) and (\ref{eq:15}--\ref{eq:20})). The errors given also indicate that, in the low-power regime, with either model, we can always find the best fittings with small errors.

Even though the two models both give sufficiently accurate simulations in the low-power regime, the model with one single photothermal timescale loses reliability when the cavity is driven at high power (\SI{2.75}{\watt} input laser as the example in FIG.~\ref{fig:3}(d)). The inconsistent result is obtained due to the deficiency of considering the thermal expansion and thermo-optic as one effect. This simpler model can still simulate the parametric amplification of random thermal fluctuations of the acoustic mode. However, the duration of these self-sustaining oscillations is not compatible with experimental observations. In contrast, the model with two photothermal effects gives a more faithful simulation by extending the duration of the oscillatory stage to the same length as the experimental data, as shown in FIG.~\ref{fig:3}(e). Furthermore, we observe that the build-up process on the left-hand side is also slowed and in closer agreement to the measured phenomenon. The amplified parametric oscillation of the simulation fits better with the experimental result on the time length of the oscillation and on the evolution of the amplitude of the oscillation.\par

Some smaller features of the simulation are still not perfectly harmonized to the experimental data. One of these is the slow power build-up on the left-hand side of resonance, which may be caused by the nonlinearity of the power dependence of the two thermal effects and the interaction between them, requiring more detailed study and analysis in the roles of these two effects playing in the parametric oscillatory instability of the system.
Other effects may need to be included for a more precise model, such as the mode restructuring in the cavity due to the shape change of the mirror resulting from its local and unequal expansion, the bolometric interaction directly coupling photothermal absorption to the acoustic mode, and the dependence of cavity decay rate and input coupling rate on either acoustic or photothermal displacements.\par

Other laser-surface interactions are not discussed in our model that may also change the mirror coating and thus the intracavity field. For example, photochemical effects usually lead to laser-induced desorption and ablation, photoacoustic effects will affect surface acoustic waves and further thermal diffusion in the materials, and repulsion force on the surface may result in the emission of neutrals (via antibonding states), ions, explosive ablation, and shockwave generation. The expansion considered in the model is the simplest effect. Other possible phenomena may respond to thermal expansion, which are commonly discussed in materials, such as deformation, stress, evaporation, and again shockwave generation are not included.\par

\section{\label{sec:level5} Conclusions}

In this work, we consider a linear optomechanical levitation system where a milligram-scale free-standing mirror acts as the oscillator. The system experiences strong nonlinearity from the photothermal effects induced by the mirror absorption of the intracavity field. The previous model with one photothermal timescale can give a faithful fitting when the input power of the system is low. With high input power, however, the model loses reliability. Here, we report an effective theoretical model, based on a thorough investigation of photothermal effects in the system. With the simulation of the thermal conduction inside the mirror coating, we separate the photothermal effects into two independent terms --- photothermal expansion and thermal-optic effects, which have different timescales. The dynamic equations of the system of this refined model successfully simulate the system at both low and high power regimes. In the high power regime, in particular, the model with two photothermal effects gives a more faithful simulation than the previous model. The higher effectiveness is observed not only in the evident extension of oscillation duration to the same length as the experimental data but also in slowing the speed of the building-up process. This sets essential groundwork for the characterization and stabilisation of existing systems and a deeper understanding of photothermal effects in optomechanical systems. The method used in the investigation of photothermal effects in this system can be easily applied to other similar optomechanical systems where high-intensity laser is used or high precision of the measurement is desired. This work facilitates the development of new and more precise optomechanical systems for integrated photonics and sensing, and paves the way to fundamental studies in quantum mechanics.\par


\section{Acknowledgements}
This work is supported by the Australian Research Council Centre of Excellence for Quantum Computation and Communication Technology (CE170100012); P.K.L. acknowledges support from the Australian Research Council Laureate Fellowship (FL150100019); G.G. acknowledges support from the Australian Research Council Discovery Program (DP230101940). We are grateful for the comments from Tao Wang.\par

\section{Data availability statement}
The data that support the findings of this study are available upon reasonable request from the authors.\par

\appendix

\section{\label{sec:A1}Two Photothermal Effects}

\begin{figure}\centering
    \includegraphics[width=0.6\textwidth]{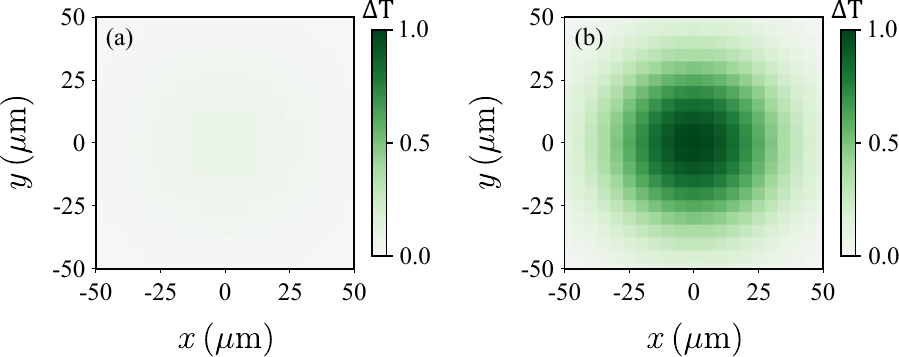}
    \caption{\label{fig:appendix}Simulation of the temperature changes on a cross-section of the mirror's surface. The temperature changes are normalized to the maximum temperature change (at the centre) at the equilibrium state. (a) gives the temperature change of the surface at \SI{0.2}{\micro\second}. (b) gives the temperature change of the surface after it has reached the equilibrium state. The simulation is under an intracavity power of \SI{300}{\watt} and an optimistic mirror absorption of $\mathcal{A}=10$ \si{ppm}. The color bar goes from lighter to darker green with the temperature change becoming larger.}
\end{figure}

The rough simulation of the thermal conduction inside the mirror coating shows that the temperature distribution is not uniform in the cross-section (see in FIG.~\ref{fig:appendix}). Here, in our assumption, we only consider the expansion and refractive index change close to the center of the laser spot. Moreover, because the thermal energy diffused to the edge of the mirror is negligible, we assume that the mirror expansion is towards the interior of the cavity. To further simplify the model, we ignore nonlinear effect on the intracavity field that results from the uneven distribution of the thermal energy and thus the reshaping of the coating surface.\par

Equations ~(\ref{eq:8}) and~(\ref{eq:9}) give two different perspectives on the effective optical length difference between cold and hot cavity. Physically, as shown in Eq.~(\ref{eq:8}), the change comes from two different components of the effective optical length. The first term represents the reduction of the distance between the input and output mirror due to the expansion of the mirror coating, whereas, the second term gives the variation of the effective optical length inside the levitation mirror.\par

However, writing in Eq.~(\ref{eq:9}) gives a mathematical perspective that is convenient for us to separate the photothermal expansion effect and the thermo-optic effect, as well as independently analyse them. The first term indicates the photothermal expansion effect on the coating part (\uppercase\expandafter{\romannumeral1}) :
\begin{eqnarray}
    \label{eq:A1}
    x_\textrm{ex\uppercase\expandafter{\romannumeral1}}(D_\textrm{\uppercase\expandafter{\romannumeral1}}(T))&=&-\left(n_\textrm{A}-n\right) \delta D_\textrm{\uppercase\expandafter{\romannumeral1}}\nonumber \\
    &=&-\left(n_\textrm{A}-n\right)\frac{\mathrm{d} D_\textrm{\uppercase\expandafter{\romannumeral1}}}{\mathrm{d}T}\delta T.
\end{eqnarray}\par

Because $D_\textrm{\uppercase\expandafter{\romannumeral1}}=D_\textrm{\uppercase\expandafter{\romannumeral1}, cold}+\delta D_\textrm{\uppercase\expandafter{\romannumeral1}}$, where $D_\textrm{\uppercase\expandafter{\romannumeral1}, cold}$ is the original (constant) length before the coating absorbing heat energy, so the time derivative of $x_\textrm{ex\uppercase\expandafter{\romannumeral1}}$ can be written as:
\begin{eqnarray}
    \label{eq:A2}
    \dot{x}_\textrm{ex\uppercase\expandafter{\romannumeral1}}&=&-\frac{\mathrm{d}(\delta D_\textrm{\uppercase\expandafter{\romannumeral1}}(T(\varepsilon ,t)))}{\mathrm{d}t}\left(n_\textrm{A}-n\right)\nonumber\\
    &=&-\frac{\mathrm{d}D_\textrm{\uppercase\expandafter{\romannumeral1}}(T(\varepsilon ,t))}{\mathrm{d}t}\left(n_\textrm{A}-n\right)\nonumber\\
    &=&-\frac{\mathrm{d}D_\textrm{\uppercase\expandafter{\romannumeral1}}}{\mathrm{d}T}\frac{\mathrm{d}T(\varepsilon ,t)}{\mathrm{d}t}\left(n_\textrm{A}-n\right)\nonumber\\
    &=&-\frac{\mathrm{d}D_\textrm{\uppercase\expandafter{\romannumeral1}}}{\mathrm{d}T}\left(\frac{\partial T}{\partial t}+\frac{\partial T}{\partial\varepsilon}\frac{\mathrm{d}\varepsilon}{\mathrm{d}t}\right)\left(n_\textrm{A}-n\right)\nonumber\\
    &=&-\frac{\mathrm{d}D_\textrm{\uppercase\expandafter{\romannumeral1}}}{\mathrm{d}T}\left(\gamma_\textrm{ex}\left(T_\textrm{0}-T\right)+\frac{\partial T}{\partial\varepsilon}P_\textrm{opt}\right)\left(n_\textrm{A}-n\right).\nonumber \\
    &&    
\end{eqnarray}
Here we assume that the mirror's expansion and the refractive index of the coating materials only depend on the temperature $T$, which changes due to the thermal energy absorption and the cooling by the environment with time. Note that the average of the dielectric refractive indices $n$ will be greater than the intracavity refractive index $n_\textrm{A}$. The thermal relaxation rate is denoted as $\gamma_\textrm{ex}$. The intracavity power is denoted by $P_\textrm{opt}$, with $\varepsilon$ which is the power absorbed by the mirror. Thus we define $\eta_\textrm{ex}$ as the ratio of the thermal energy that the mirror absorbs from the intracavity field $\eta_\textrm{ex} = \mathrm{d}P_\textrm{opt}/\mathrm{d}\varepsilon$. It is proportional to the reciprocal of the thermal absorption coefficient. \par

We then further simplify the expression, by replacing $\left(T-T_\textrm{0}\right)$ with $\delta T$, and $\partial T/\partial \varepsilon$ with $\left(\mathrm{d}T/\mathrm{d}P_\textrm{opt}\right)\left(\mathrm{d}P_\textrm{opt}/\mathrm{d}\varepsilon\right)$, we have
\begin{eqnarray}
    \label{eq:A3}
    \dot{x}_\textrm{ex\uppercase\expandafter{\romannumeral1}}&=&-\frac{\mathrm{d}D_\textrm{\uppercase\expandafter{\romannumeral1}}}{\mathrm{d}T}\left(-\gamma_\textrm{ex}\delta T+\eta_\textrm{ex}\frac{\mathrm{d} T}{\mathrm{d}P_\textrm{opt}}P_\textrm{opt}\right)\left(n_\textrm{A}-n\right)\nonumber\\
    &=&-\gamma_\textrm{ex}\Big[-\left(n_\textrm{A}-n\right)\frac{\mathrm{d}D_\textrm{\uppercase\expandafter{\romannumeral1}}}{\mathrm{d}T} \delta T\nonumber \\
    &&+\left(\frac{\eta_\textrm{ex}}{\gamma_\textrm{ex}}\left(n_\textrm{A}-n\right)\frac{\mathrm{d}D_\textrm{\uppercase\expandafter{\romannumeral1}}} {\mathrm{d}T}\frac{\mathrm{d}T}{\mathrm{d}P_\textrm{opt}}\right)P_\textrm{opt}\Big].
\end{eqnarray}
Substituting $x_\textrm{ex\uppercase\expandafter{\romannumeral1}}$ with Eq.~(\ref{eq:A1}) into Eq.~(\ref{eq:A3}),
\begin{eqnarray}
    \label{eq:A3_2}
    \dot{x}_\textrm{ex\uppercase\expandafter{\romannumeral1}}
    &=&-\gamma_\textrm{ex}\left(x_\textrm{ex\uppercase\expandafter{\romannumeral1}}+\frac{\partial D_\textrm{\uppercase\expandafter{\romannumeral1}}}{\partial P_\textrm{opt}}P_\textrm{opt}\right)\nonumber\\
    &=&-\gamma_\textrm{ex}\left(x_\textrm{ex\uppercase\expandafter{\romannumeral1}}+\beta_\textrm{ex\uppercase\expandafter{\romannumeral1}}P_\textrm{opt}\right),
\end{eqnarray}
where $\beta_\textrm{ex\uppercase\expandafter{\romannumeral1}}$ is susceptivity coefficient of photothermal expansion. Eq.~(\ref{eq:A3_2}) gives a clear idea of how the expansion of the coating part (\uppercase\expandafter{\romannumeral1}) affected by the intracavity power (the second term) and the decay with time (the first term). The intracavity field heat the coating up, causing expansion on it and thus leading to a decrease of the cavity length. By replacing $(n_\textrm{A}-n)$ with $n$ and $D_\textrm{\uppercase\expandafter{\romannumeral1}}$ with $D_\textrm{\uppercase\expandafter{\romannumeral2}}$, combining Eq.~(\ref{eq:11}), an analogous calculation gives the derivative of the effective optical length change of part (\uppercase\expandafter{\romannumeral2}):
\begin{eqnarray}
    \label{eq:A4}
    \dot{x}_\textrm{ex\uppercase\expandafter{\romannumeral2}}=-\gamma_\textrm{ex}(x_\textrm{ex\uppercase\expandafter{\romannumeral2}}+\beta_\textrm{ex\uppercase\expandafter{\romannumeral2}}P_\textrm{opt}),
\end{eqnarray}
where $\beta_\textrm{ex\uppercase\expandafter{\romannumeral2}}=\partial D_\textrm{\uppercase\expandafter{\romannumeral2}}/\partial P_\textrm{opt}$.\par

Because the photothermal expansion effect in part (\uppercase\expandafter{\romannumeral1}) and (\uppercase\expandafter{\romannumeral2}) share the same values of $\gamma_\textrm{ex}$, we can simply add up the two equation~(\ref{eq:A3}) and (\ref{eq:A4}) to get the time derivative of the thermal expansion in the whole mirror coating:
\begin{eqnarray}
    \label{eq:A6}
    &\dot{x}_\textrm{ex}=-\gamma_\textrm{ex}(x_\textrm{ex}+\beta_\textrm{ex}P_\textrm{opt}),
\end{eqnarray}
where $\dot{x}_\textrm{ex}=\dot{x}_\textrm{ex\uppercase\expandafter{\romannumeral1}}+\dot{x}_\textrm{ex\uppercase\expandafter{\romannumeral2}}$, $x_\textrm{ex}=x_\textrm{ex\uppercase\expandafter{\romannumeral1}}+x_\textrm{ex\uppercase\expandafter{\romannumeral2}}$ and $\beta_\textrm{ex}=\beta_\textrm{ex\uppercase\expandafter{\romannumeral1}}+\beta_\textrm{ex\uppercase\expandafter{\romannumeral2}}=\partial D/\partial P_\textrm{opt}$ if susceptivity coefficient of part (\uppercase\expandafter{\romannumeral1}) and (\uppercase\expandafter{\romannumeral2}) are independent to each other. This equation shows the power dependence of the expansion, where a positive susceptivity coefficient represents an expansion of the materials after absorbing the thermal energy, giving a reduction of the effective optical length.\par

Similarly, we can get the derivative of the effective optical length variation of thermo-optic effect as
\begin{eqnarray}
    \label{eq:A7}
    x_\textrm{re}(n(T))&=&D_\textrm{\uppercase\expandafter{\romannumeral1}}\frac{\mathrm{d} n}{\mathrm{d} T}\delta T,
\end{eqnarray}
and then we have the time derivative of it: 
\begin{eqnarray}
    \label{Eq:thermo_optic_dot_x}
    \dot{x}_\textrm{re}&=&\frac{\mathrm{d}n(T(\varepsilon ,t))}{\mathrm{d}t}D_\textrm{\uppercase\expandafter{\romannumeral1}}\nonumber\\
    &=&\frac{\mathrm{d}n}{\mathrm{d}T}\frac{\mathrm{d}T(\varepsilon ,t)}{\mathrm{d}t}D_\textrm{\uppercase\expandafter{\romannumeral1}}\nonumber\\
    &=&\frac{\mathrm{d}n}{\mathrm{d}T}\left(\frac{\partial T}{\partial t}+\frac{\partial T}{\partial\varepsilon}\frac{\mathrm{d}\varepsilon}{\mathrm{d}t}\right)D_\textrm{\uppercase\expandafter{\romannumeral1}}\nonumber\\
    &=&\frac{\mathrm{d}n}{\mathrm{d}T}\left[\gamma_\textrm{re}\left(T_\textrm{0}-T\right)+\frac{\partial T}{\partial\varepsilon}P_\textrm{opt}\right]D_\textrm{\uppercase\expandafter{\romannumeral1}}\nonumber\\
    &=&-\gamma_\textrm{re}\left(x_\textrm{re}-\frac{\partial n}{\partial P_\textrm{opt}}P_\textrm{opt}\right)\nonumber\\
    &=&-\gamma_\textrm{re}\left(x_\textrm{re}-\beta_\textrm{re}P_\textrm{opt}\right)
\end{eqnarray}
where $\gamma_\textrm{re}$ and $\beta_\textrm{re}$ are thermal relaxation rate and susceptivity coefficient of thermo-optic effects. \par

Here we give the detailed derivation to emphasise the different sign before susceptivity coefficient compared with the photothermal expansion (Eq.~\ref{eq:A6}), which indicates that with the positive $\beta_\textrm{re}$, the refractive index increases when absorbing thermal energy, reducing the deduction of effective optical length. Whereas, in our system, with a positive $\beta_\textrm{re}$, the refractive index decreases after absorbing thermal energy, leading to a further decrease of the effective optical length.\par

\end{document}